\documentclass[graybox]{svmult}

\usepackage{mathptmx}       
\usepackage{helvet}             
\usepackage{courier}            
\usepackage{type1cm}         
\usepackage{makeidx}         
\usepackage{graphicx}        
\usepackage{multicol}        
\usepackage[bottom]{footmisc}
\usepackage{natbib}
\usepackage{amsbsy}
\newcommand{\teff}{\ensuremath{T_{\mathrm{eff}}}}
\newcommand{\logg}{\ensuremath{\log g}}
\newcommand{\feh}{\ensuremath{[\rm Fe/\rm H]}}

\newcommand{\qurad}{Q_{\rm rad}}
\newcommand{\frad}{F_{\rm rad}}
\newcommand{\quvis}{Q_{\rm visc}}

\newcommand{\nab}{\nabla}

\makeindex     

\begin{document}

\title*{Analysis of stellar spectra with 3D and NLTE models}
%
%
\author{Maria Bergemann}
%
%
\institute{M. Bergemann \at Institute  of Astronomy, University  of Cambridge, 
Madingley Road, CB3  0HA, Cambridge, UK \email{mbergema@ast.cam.ac.uk}}%
%
%
\maketitle

\abstract*{}
\abstract{Models of radiation transport in stellar atmospheres are the hinge of
modern astrophysics. Our knowledge of stars, stellar populations, and galaxies
is only as good as the theoretical models, which are used for the interpretation
of their observed spectra, photometric magnitudes, and spectral energy
distributions. I describe recent advances in the field of stellar atmosphere
modelling for late-type stars. Various aspects of radiation transport with 1D
hydrostatic, LTE, NLTE, and 3D radiative-hydrodynamical models are briefly
reviewed.}

\section{Introduction}
\label{sec:1}
Models of stellar atmospheres and spectral line formation are a crucial part
of observational astrophysics. The models are our ultimate link between
observations of stars and their fundamental physical parameters. On the one
side, the models allow us to go from observable quantities, i.e stellar fluxes,
spectral energy distributions, and photometric
magnitudes, to physical parameters of stars, such as effective temperature
$\teff$, surface gravity $\logg$, metallicity $\feh$, abundances, rotation and
turbulent velocities. The atmospheric models also allow us to convert
theoretical bolometric luminosities from stellar evolution models to their
observable quantities, theoretical colours, which are then compared with
observations. E.g., model vs observed colour-magnitude diagrams are used to
determine distances and ages of clusters and field stars. The shapes of 
stellar energy distributions (SED) serve as a diagnostics of inter-stellar  and
circumstellar reddening. There are mass- and age-sensitive diagnostics in
stellar spectrum, such as the ultra-violet Ca H and K lines, which may help to
tell a
young star from an old star. Detailed chemical abundances are arguably the most
important physical quantities, which can be only deciphered from a stellar
spectrum. They link stellar properties to nucleosynthesis of elements in the Big
Bang, in stars at the end of their life-times, in violent explosions, caused by
stellar interactions, and by cosmic ray acceleration in the interstellar medium.
This connection forms the basis of important diagnostics methods
to constrain formation and evolution of stars, stellar populations, and
galaxies.

In short, all fundamental physical stellar quantities critically depend on the
models we use for the analysis of observations.  Until recently, we could only
use the simplest 1D hydrostatic models in local thermodynamic equilibrium (LTE)
for stellar parameter determinations, for more sophisticated models were far
too computationally expensive. Calculation of a 3D radiative-hydrodynamical (3D
RHD) model required months of CPU time, which was of little use for any
practical application in spectroscopic analysis. Moreover, the accurate solution
of radiation transport even in the time-independent scheme has been too
demanding for quantitative applications.  Non-local thermodynamic equilibrium
(NLTE) calculations were disfavoured for several reasons. First, inversion of
large matrices in the complete linearisation scheme (for solving coupled
statistical equilibrium equations in non-LTE) was prohibitive for atoms with
complex atomic structure.  Furthermore, lack of accurate ab initio
calculations or experimental atomic data lead to the need to introduce
simplified treatment of several types of atomic processes in vastly more complex
non-LTE calculations, which caused a wide-spread misconception that non-LTE
calculations are of questionable advantage.

\begin{svgraybox}
It is crucial to realise that non-LTE is not an approximation, in contrast to
LTE, which approximates all collision rates by the infinitely large numbers, and
fully ignores the influence of radiation field in stellar atmospheres on the
energy distribution of matter. One may however, solve time-dependent rate
equations, or drop the time dependence, which reduces the problem to solving the
equations of statistical equilibrium only (see below).
\end{svgraybox}

However, the theory of radiative transfer in stellar atmospheres is one of the
most mature fields in astrophysics (e.g. the fundamental work by
\citealt{Mihalas:1979ux}) and the numerical implementation of the theory has
seen substantial improvement over the past decade. Moreover, we have enough
computational power to simulate stellar convection in 3D and solve for radiative
transfer explicitly taking into account the interaction of gas particles with
the radiation field. In this lecture I summarise the main progress in modelling
atmospheres and spectra of cool, FGKM, stars, that has been made during the past
decade, and provide a timeline for the developments in the field which can be
expected in the near future. I do not touch upon the problems of radiative
transfer in more complex cases, such as expanding supernova shells
\citep{Lucy:1999wd} or  dusty AGB envelopes \citep[][ and ref.
therein]{Hofner:2003vf}, or chromospheres \citep[e.g.][]{Hansteen:2007wn}. 

\section{Basic considerations}

The analysis of cool FGK stars meets with two main kinds of difficulties. These
stars have sub-surface convection zone and radiation field in their photospheres
is highly non-local. As a consequence, neither of the two classical modelling
approximations, LTE and 1D hydrostatic equilibrium, is valid and cannot be
justified without detailed \textit{ab initio} theoretical calculations. An
excellent review of NLTE and 3D simulations from a theoretical perspective is
given in the Annual Review articles by \cite{Mihalas:1973bu} and
\cite{Spruit:1990cv}; a comprehensive summary of 3D and NLTE in application to
stellar abundance analysis is provided in \cite{2005ARA&A..43..481A}.

What is the reason for NLTE? A star is not a perfect black-body and radiation
field influences the physical state of matter and vice versa. Therefore, only
consistent NLTE radiative transfer will correctly recover the thermodynamic
structure of the atmosphere and thus the properties of its radiation field. In
NLTE, the particle distribution functions, i.e. the excitation and ionisation
states, are influenced by the radiation field that causes them to depart from
the LTE values. Why this happens is easy to understand. Stellar atmosphere is
the region where radiation escapes into ISM. Deep in the atmospheres, opacity is
large and the photon mean free path is very small. Radiation transport occurs
over very small length scales, thus connecting gas parcels with very similar
thermodynamic properties. 
\begin{svgraybox}
The definition of a \textit{stellar surface} is somewhat vague. Strictly
speaking, this is the boundary where material at the most transparent wavelength
$\lambda$ becomes optically thin, $\tau_\lambda < 1$. This wavelength is $1.6
\mu$m, where the $H^{-}$ opacity in continuum is at its minimum, however, in
practice this wavelength is rarely used. It is more common to refer to the
$\lambda = 500$ nm, where most of the flux is emitted by a typical solar-like
star, or to the Rosseland mean opacity, a Planck-function weighted average over
all frequencies.

Photon mean free path is the length scale over which thermalisation takes place.
\end{svgraybox}
However, closer to the stellar surface photon mean free path $\lambda_p$ becomes
large, larger than the scale height of material. Thus, as we move outwards, the
decoupling between radiation and matter increases, radiation becomes non-local,
anisotropic, and strongly non-Planckian. Many spectral lines form in NLTE
showing shapes which bear no resemblance with the line profiles predicted by
classical LTE models. For example, the Mg I line at 12 $\mu$m reverts to
emission at the solar limb that can be only accounted for by NLTE
models. The effect on other spectral features is less obvious, although it may
still dramatically impact the line equivalent widths and their detailed
profiles, i.e. the total energy absorbed in a line (see next Section).

\begin{figure}[!ht]
\begin{center}
\includegraphics[scale=.5]{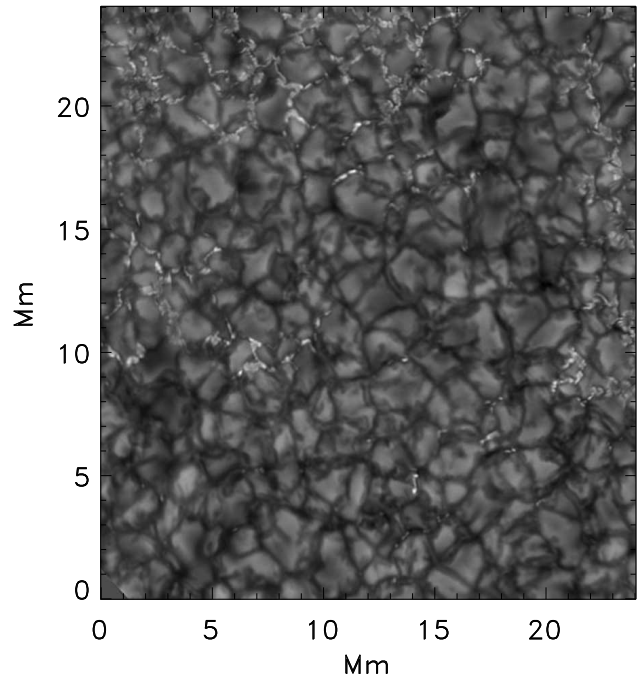}
\caption{Solar granulation in the G-continuum \citep{Nordlund:2009jf}.
Observations were made on the Swedish 1m Solar Telescope (Institute of
Theoretical Astrophysics, Oslo). Reproduced by permission from the authors and
the publisher.}
\label{fig:one}
\end{center}
\end{figure}

\begin{figure}[!ht]
\begin{center}
\includegraphics[scale=.5]{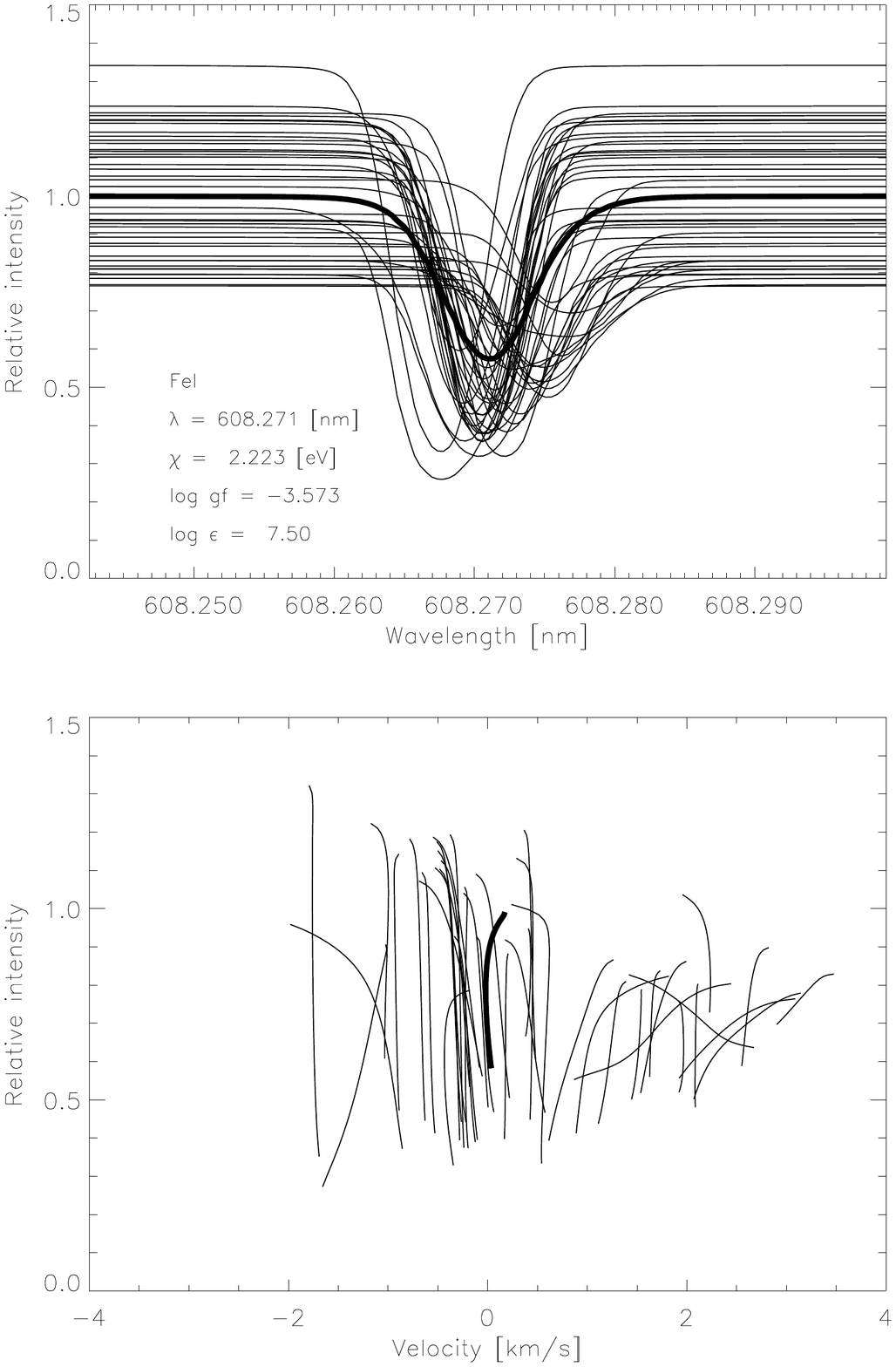}
\caption{Line profiles of the Fe I line at 6082 $\AA$ in the 3D RHD simulations
of the solar 
atmosphere \citep{2000A&A...359..729A}. Top panel: spatially-resolved profiles.
Bottom panel: the line bisectors. The thick lines on both panels indicate the
spatially-averaged profile and the $C$-shaped bisector. The line asymmetries are
a consequence of blue- and red-shifts caused by the influence of the convective
velocity fields on the line formation. Reproduced by permission from the authors
and the publisher.}
\label{fig:two}
\end{center}
\end{figure}

Another major source of complexity is convection. There are several observable
manifestations of stellar convection, the most prominent is stellar surface
granulation, which has been in the focus of observational research for more than
a decade. Fig. \ref{fig:one} shows a patch on the solar surface observed with
the Swedish Solar Telescope telescope. The intensity fluctuations correspond to
brighter granules and darker inter-granular lanes. The brightness fluctuations
are about $20\%$, while the actual temperature contrast across the granulation
pattern is about $25\%$. This mild dependence of surface intensity variation on
$T$ is caused by the huge sensitivity of H$^-$ continuum opacity to the
temperature. The other key observable are the characteristic $C$- and inverse
$C$- shaped bisectors of the line profiles that are caused by up- and
down-flows \citep[e.g.][]{2008AJ....135.2033G}. The bisectors can be best
observed in the very high-quality solar spectrum (Fig. \ref{fig:two}).

Indeed, these beautiful observations motivated the development of complex 3D
radiative hydrodynamics (RHD) models \citep{Nordlund:2009jf} and even larger
solar telescopes, such as, e.g., the new German 1.5m GREGOR telescope.
Inhomogeneities, spots, and convective motions were also resolved on the
surfaces of red supergiants \citep[][also see the first lecture in this
series]{1990MNRAS.245P...7B, 1992MNRAS.257..369W, 1997MNRAS.285..529T,
1997MNRAS.291..819W, 2000MNRAS.315..635Y, 2009A&A...508..923H,
2009A&A...503..183O, Kervella:2009di, 2011A&A...529A.163O}. More interesting
results about atmospheric structure and dynamics of cool stars will soon come to
light with new telescopes and missions, such as the Herschel Space Observatory
\citep{Teyssier:2012bz}. All these data will need complex theoretical models to
understand the physics underlying the observed phenomena.

\section{Theory}

What we are trying to understand is how the stellar light we detect with our
instruments, e.g. in the form of a spectrum, is created in a stellar atmosphere.
Thus, the essence of problem is to solve the radiative transfer equation, which
describes the temporal and spatial evolution of the radiation field $J_\nu$:

\begin{equation}  \label{eq:radtran} 
1/c \frac{dI_\nu}{dt} + \frac{dI_\nu}{ds} = \eta_\nu  - \alpha_\nu I_\nu 
\end{equation}

where $ds$ is geometrical path length along the beam, $\eta_\nu$ and
$\alpha_\nu$ are the monochromatic (linear or volume) emission and extinction
coefficients. Clearly, the transport of radiation depends on the medium in which
it propagates. Thus, we also need a model that describes the thermodynamic
properties in stellar atmospheres, which must include radiation transport. 

There are several types of models, vastly different in complexity and
computational burden: from the simplest 1D LTE hydrostatic models to the
(presently) most sophisticated 3D RHD LTE models with scattering. In
all cases, the radiative transfer is assumed to be \textit{quasi-static}, i.e.
the time derivative is neglected, which is not a bad approximation because the
photon propagation time is much shorter than the fluid motion time. The
computational timescales range from few seconds of CPU to several months.  In
the next section, I will summarise the main ingredients of all these model types
without going into details of how the equations are solved (see e.g.
\citealt{Mihalas:1979ux}).

\subsection{Classical 1D static model atmospheres}

The simplest models are computed using the 1D hydrostatic LTE approximations. We
have to solve several equations: 

\begin{enumerate}
\item equation of radiative transfer in the static, time-independent, case. For
a plane-parallel atmosphere with a geometrical depth $z$ (it is straightforward
to generalise this to a spherically-symmetric case):

\begin{equation}  \label{eq:radtran} \cos \theta \frac{dI_\nu}{dz} = \alpha_\nu
I_\nu - \eta_\nu \end{equation}
The equation describes a change of specific intensity $I_\nu$, as radiation
passes through a slab of a thickness $dz$. For the plane-parallel approximation,
the latter is represented by the projection at angle $\theta$ between the
direction of the light beam and the $z$ axis. The parameters $\alpha_\nu$ and
$\eta_\nu$ with dimensions cm$^{-1}$ are the monochromatic linear extinction and
emission coefficients; their ratio gives a source function $S_\nu =
\frac{\eta_\nu}{\alpha_\nu}$. In LTE calculations, the source function is equal
to the Planck function $S_\nu = B_{\nu}$ if only true absorption and emission
processes are considered (strict LTE), or approximated by a two-level form, 
$S_\nu = (\kappa_\nu B_{\nu} + \sigma_\nu J_\nu)/(\kappa_\nu + \sigma_\nu)$,
where $\kappa$ is the true absorption and $\sigma$ the scattering coefficients. 
In 'less strict' LTE, one may include, e.g. a coherent isotropic scattering in
continuum, e.g. Thomson scattering on electrons $e^-$.
\item equation of flux constancy, or energy conservation:

 \begin{equation} 
 F= \frac{L}{4 \pi R^2} = \sigma \teff^4,
 \end{equation}
where $F$ is the bolometric flux, $L$ the luminosity, $R$ the stellar radius,
and $\teff$ the effective temperature. In other words the divergence of the
total energy flux transported to the surface is zero, $\nab F = 0$ (see the next
section on 3D RHD models), and the total flux is equated to the integral flux
from the surface of a black body. 

The total flux is usually taken to be the sum of convective and radiative
components, $F = F_{conv} + F_{rad}$. The convective flux, $F_{conv}$, is needed
because in deeper atmospheric layers, energy is mostly carried by convection. In
the  standard 'mixing-length' type approximations, $F_{conv} \sim
\frac{\alpha_{MLT}}{H_p}$, where $H_p$ is pressure scale height and
$\alpha_{MLT}$ the mixing-length parameter. There are attempts to calibrate the
mixing length parameter based on 3D models (see next Section).

\item radiative equilibrium equation\footnote{Assuming that energy is
transported by radiation only.}:
\begin{equation}  
\label{eq:radeq} 
\int_{0}^{\infty} \alpha_\nu J_{\nu}(\tau)d\nu =
\int_{0}^{\infty} \alpha_\nu S_{\nu}(\tau)d\nu 
\end{equation}
where $\tau$ is the optical depth, $J_\nu$ the mean intensity averaged over all
directions,  and $S_\nu$ the source function. The left-hand side of the equation
can be viewed as radiation 'sink' term, and right-hand side is the radiation
'source' term, which in LTE is usually equal to the Planck function  (see the
next section for how this is treated in 3D RHD models). In the calculations of
energy balance with 1D hydrostatic models, it is presently possible to evaluate
detailed opacity on more than $10^5$ wavelength points
\citep[e.g.][]{Gustafsson:2008df, Grupp:2004jc}. This samples the majority of
the atomic and molecular lines, which contribute to the opacity under conditions
typical of a late-type star.
\item equation of hydrostatic equilibrium:
\begin{equation}  
\label{eq:hydeq} 
\nab P_{tot} = - \rho \frac{G M_r}{r^2},
\end{equation}
where $\rho$ is the density, $M_r$ mass at the radius $r$, $\nab P_{tot} = \nab
P_{gas} + \nab P_{rad}$, i.e the sum of gas and radiative pressures. The latter
can be expressed as:

\begin{equation}  
\label{eq:radpres} 
\nab P_{rad} = -  1/c \int_{0}^{\infty}(\kappa_\nu + \sigma_\nu)
F_\nu d\nu
\end{equation}

Gas pressure is balanced by the gravitational attraction, which does not change
with depth in plane-parallel models. In spherically-symmetric models, $g=g(r)$,
thus mass (M) and luminosity ($L = 4 \pi r^2 F(r)$) are constant throughout the
atmosphere. Note that there are some variations, e.g. some models include a
turbulent pressure term, $ \nab P_{turb} \sim \rho \upsilon_t^2$, where
$\upsilon_t$ is the characteristic velocity, which may be used e. g. in very
extended atmospheres (red giant branch or red supergiant stars) to approximate
the levitation of the photosphere, the effect of 3D hydrodynamics. For detailed
numerical schemes and practical implementation of the equations, the reader is
referred to  \cite{Kurucz:1993wc}, \cite{Grupp:2004jc},
\cite{Gustafsson:2008df}.

\end{enumerate}

\begin{figure}[ht]
\begin{center}
\includegraphics[scale=.65]{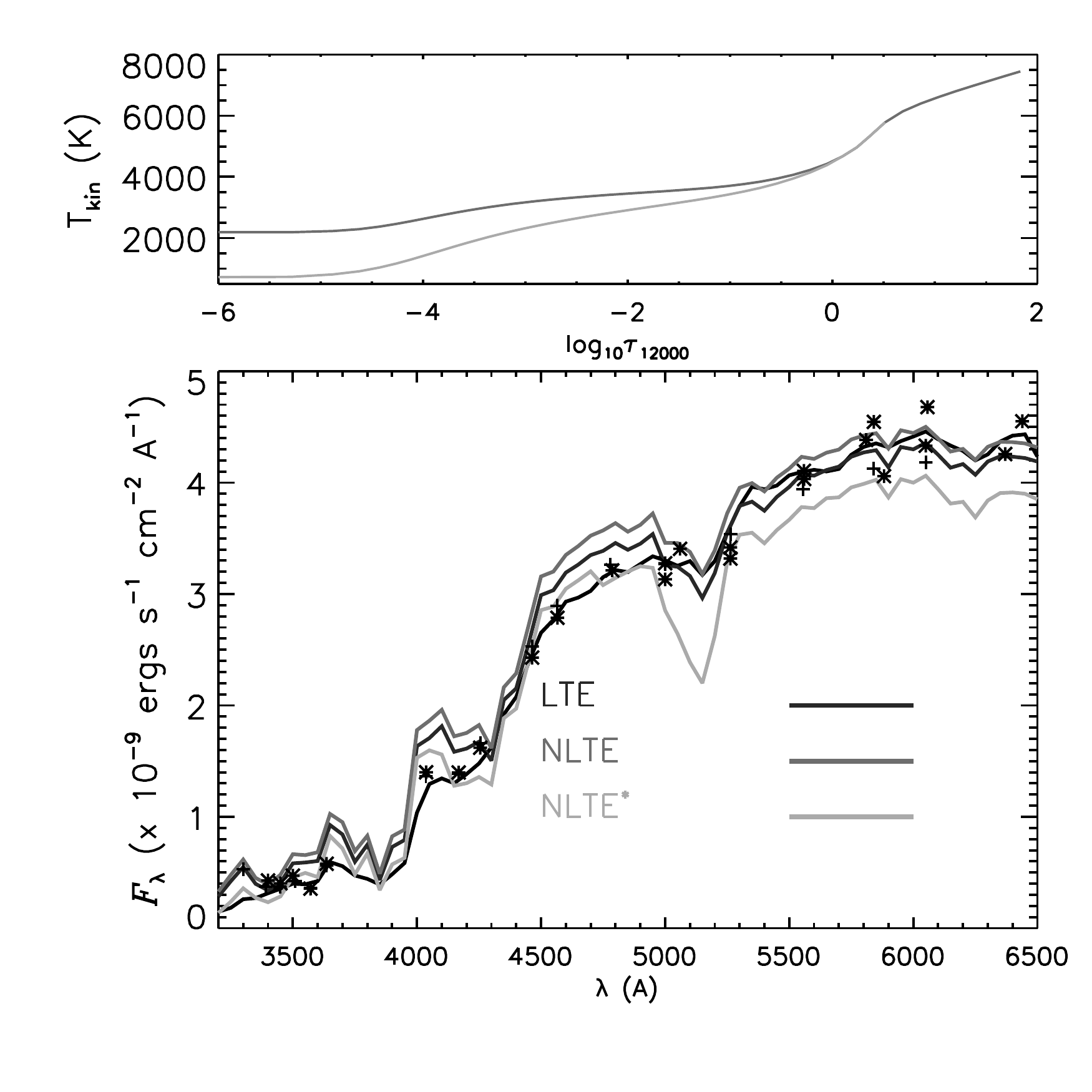}
\caption{NLTE model atmospheres (top) and synthetic spectra (bottom) for
Arcturus, a very cool RGB star (dark - LTE, light - NLTE). NLTE$^*$ refers to
the NLTE calculations with an ad-hoc kinetic temperature structure adjusted to
fit the near-UV SED; note however that the fit in the optical has degraded.
Reproduced by permission from the authors \citep{Short:2009iq} and from the
publisher.}
\label{fig:phoenix}
\end{center}
\end{figure}

With the 1D hydrostatic approximation, it is also possible to compute
fully-consistent NLTE models including the feedback on the atmospheric
structure. Such models can be computed with the PHOENIX code
\citep{Baron:1998dn}, which is very versatile in its applicability. In NLTE,
PHOENIX includes more than 20 elements in the rate equations. NLTE model
atmospheres, because of the UV overionization in Fe, lead to \textit{warmer}
outer atmospheric layers with respect to LTE for stars with $\teff > \sim 5400$
K \citep{Short:2005kh, Short:2009iq}. However, surface cooling by $\sim 150$ K
is predicted for the cooler stars, like the RGB star Arcturus (Fig.
\ref{fig:phoenix}, top panel), which is caused by the UV over-ionisation in the
bound-free edges of Mg, Si, and Al. This has a major impact on the UV fluxes of
cool stars (Fig. \ref{fig:phoenix}, bottom panel).

\subsection{3D RHD model atmospheres}

Full time-dependent, 3D, RHD simulations of radiative and convective energy
transport in stellar atmospheres exist since more than 2 decades. The seminal
papers in this field are \citet{1982A&A...107....1N},
\citet{1998ApJ...499..914S}, and \citet{Spruit:1990cv}. 3D RHD models have
substantially gained in complexity since then (e.g.
\citealt{Nordlund:2009jf}, \citealt{Chiavassa:2011ew},  \citealt{Collet:2011ga},
\citealt{Beeck:2012hn}), in particular in the treatment of radiation transport.
More details about the state-of-the-art models and their application to
modelling stellar spectra will be given in the next section, here I only briefly
review the equations of radiative hydrodynamics,
needed to compute the models.

The main difference with respect to hydrostatic models (see above) is that one
solves time-dependent  fluid-hydrodynamics equations for compressible fluid
flow. In Eulerian form, the equations are:
\begin{enumerate}
\item  the equation of continuity or mass conservation
\begin{equation}
\frac{\partial \rho}{\partial t}  = - \nab \cdot (\rho \pmb{\upsilon}),
\end{equation}
where $\pmb{\upsilon} = \pmb{\upsilon}(x,z,y,t) $ is the velocity vector. The
equation essentially means that in the ascending flow, the rapid decrease of
density can be balanced by the rapid expansion sideways. 

\item equation of motion
\begin{equation}
\frac{\partial \rho \pmb{\upsilon}}{\partial t}  = - \nab \cdot (\rho
\pmb{\upsilon\upsilon}) -
\nab P - \rho \nab \Phi - \nab \cdot \tau_{visc},
\end{equation}
where $P$ is the gas pressure, $\Phi$ the gravitational potential, $\tau_{visc}$
the viscous stress tensor. The first term on the right-hand side is the
divergence of the vertical component of Reynold's stress tensor, i.e. the force
density on the fluid due to the turbulent fluctuations. Under the assumption of
slow and bulky fluid motions, one simply recovers the equation of hydrostatic
equilibrium (eq. \ref{eq:hydeq}). 

In the 'box-in-a-star' simulations, the gravitational potential is a constant,
so $d \Phi/d z = -g$, surface gravity. Box-in-a-star simulations are usually
applied to modelling solar-like stars \citep[e.g.][]{1998ApJ...499..914S}.

In contrast, in the 'star-in-a-box' setup, the simulation box includes the whole
star \citep[e.g.][]{2012JCoPh.231..919F}. In the 'star-in-a-box' regime, one
assumes a spherical potential in the form:
\begin{equation}
\Phi  = -\frac{GM}{(r_0^4 + r^4/\sqrt 1+(r/r_1)^8)^{1/4}}
\end{equation}
where r$_0$ and r$_1$ are the so-called 'smoothing parameters'. They provide
limiting values of gravitational potential for $r \to 0$ and $r \to \infty$. For
large stars-in a box models, $r_0 \sim 0.2~R_{star}$. This setup
is applied to model very extended, variable and stochastically pulsating stars,
such as cool giants and red supergiants. Note that the size of granules is very
different for solar-like and giant stars, it scales with the pressure scale
height
\citep{2002AN....323..213F}.

\item energy conservation
\begin{equation}
\frac{\partial E}{\partial t} = - \nab \cdot ( \pmb{\upsilon} (E_{i}  + P)) -
\frac{1}{2} \nab \cdot \rho \upsilon^2 \pmb{\upsilon} + Q_{\rm rad} + Q_{\rm
visc}
\end{equation}
where $E$ is the total energy density, i.e. the sum of internal E$_{i}$ and
kinetic E$_{kin}$ energies, $\quvis$ is the viscous dissipation term and
$\qurad$ is the radiative term, which is expressed as:
\begin{equation}
Q_{\rm rad} = - \nab \cdot \frad = 4 \pi \int_{0}^{\infty} \alpha_\nu
(I_\nu - S_\nu)
d\Omega d\nu
\end{equation}
here the intensity $I$ comes from the solution of radiation transport equation,
which is done separately on a system of rays for different outgoing directions.
Radiative transfer calculations are very expensive in 3D, because it is
necessary to compute several radiative ray directions for every convective time
step in the RHD computation. Therefore the radiation field is computed using 
opacity bins, which are constructed by sorting all (line plus continuum)
opacities into groups according to their amplitude \citep{1982A&A...107....1N}:
\begin{equation}
Q_{\rm rad} \equiv \sum_{n=1,  n_{bins}} (J_{bin} - B_{bin}) w_{bin},
\end{equation}
where $J$ and $B$ are the mean intensity and the Planck function, and each
quantity in this equation refers to a 'bin' rather than to a  wavelength.
$w_{bin}$ is the weight of each bin, computed as a sum of the weights of
individual wavelength points in a bin. Note that in this definition, one
assumes that the Planck function does not vary much across a bin. Modern codes
(such as Stagger) use up to 12 opacity bins, which sort the opacities according
to magnitude and wavelength.
\end{enumerate}
There are several codes available for 3D RHD modelling: the Copenhagen Stagger
code \citep{Nordlund:1992cq}, MuRAM \citep{2005A&A...429..335V}, BIFROST
\citep{2011A&A...531A.154G}, CO5BOLD \citep{2012JCoPh.231..919F,
Chiavassa:2011ew}, ANTARES \citep{Muthsam:2010dq}. All these codes are capable
of simulating stellar convection in the box-in-a-star regime. The CO5BOLD code
also supports the star-in-a-box simulations of stellar convection.

\begin{figure}
\begin{center}
\includegraphics[scale=.35]{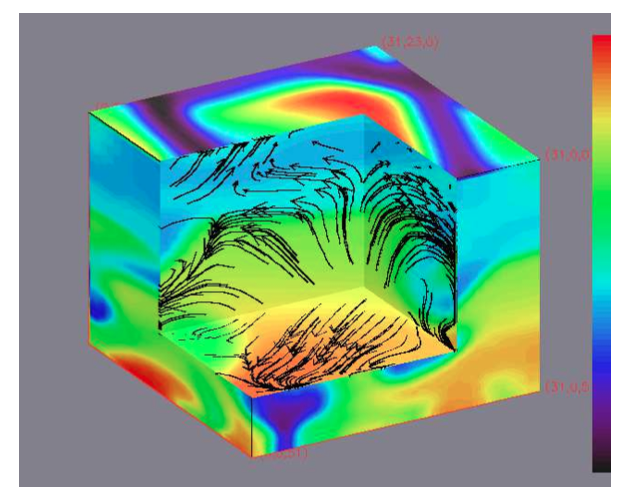}
\caption{A granule in the simulation of the solar convection. The colour bar
indicates temperature. By permission from the authors \citep{Nordlund:2009jf}
and from the publisher.
\label{fig:granule}}
\end{center}
\end{figure}

In the box-in-a-star regime, the physical domains of the simulations cover a
representative portion of the stellar surface. Vertically, they include the
whole photosphere as well as the upper part of the convective layers, typically
encompassing $12$ to $15$ pressure scale heights. Horizontally, they extend over
an area sufficiently large to host  at least about ten granules at the surface.
Fig. \ref{fig:granule} shows a snapshot from a 3D RHD simulation of the solar
surface convection. The granule characterised by warm up flowing material is
surrounded by cooler inter granular lanes, in which material sinks under the
influence of gravity. Such models predict a different temperature and pressure
structure of a stellar atmosphere, especially for stellar parameters different
from the Sun. For example, at low metallicity, 1D hydrostatic models severely
overestimate the average temperatures of the upper atmospheric layers, with
profound implications for the spectral line formation
\citep[e.g.][]{2007A&A...469..687C}.

\begin{figure}
\begin{center}
\includegraphics[scale=.45]{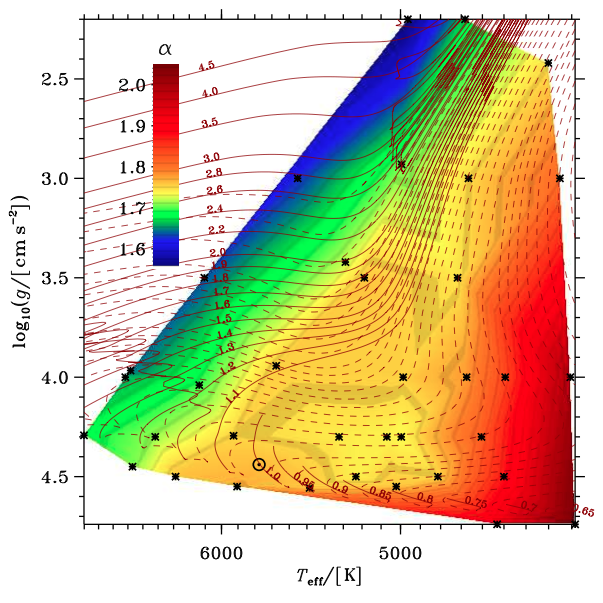}
\caption{The behavior of the calibrated MLT $\alpha$ with $T_{\rm eff}$ and
$g_{\rm surf}$ (Fig. courtesy of R. Trampedach, see also
\citealt{2013ApJ...769...18T}). The $T_{\rm eff}$-scale is logarithmic. We have
also plotted evolutionary tracks produced with the MESA-code
\citep{paxton:MESA}, covering the mass-range 0.65 - 4.5\,$M_\odot$, as
indicated. The dashed lines correspond to the pre-main-sequence evolution. The
$\alpha$-values were interpolated linearly on Thiessen triangles  between the
simulations, and indicate the values with colors as shown on the color bar. The
solar simulation is indicated with a $\odot$ and the locations of the other
simulations are shown with asterisks.
\label{mlt}}
\end{center}
\end{figure}

One may also use 3D RHD models is to devise smart calibration relationships,
which can be applied in 1D hydrostatic computations to better approximate the
complex physical processes. One of the most interesting applications is the
calibration of the \textit{mixing-length parameter}, $\alpha_{MLT}$. Fig.
\ref{mlt} shows the behaviour of $\alpha$ with the effective temperature $T_{\rm
eff}$ and surface gravity $\logg$. The $T_{\rm eff}$-scale is logarithmic.
Evolutionary tracks \citep{paxton:MESA} are also indicated. The $\alpha$-values
are indicated with colours. The solar simulation is indicated with a $\odot$ and
the locations of the other simulations are shown with asterisks. In all 1D
hydrostatic models (such as MARCS, Kurucz, MAFAGS), the mixing-length is
constant, and, depending on the formulation (e.g. \citealt{BohmVitense:1958vy},
or \citealt{Canuto:1991bj}), it takes values from $0.5$ to $1.5$. From Fig.
\ref{mlt}, it is obvious that none of these factors are physically sensible
across the full HRD. At best, a constant $\alpha$ could be used for the stars
very similar to the Sun, but not for any other star.

Grids of 3D models have been presented in \citet[][14 models]{Tanner:2013gx}, 
\citet[][37 models]{2013ApJ...769...18T},  \citet[][77
models]{2009MmSAI..80..711L}, \citet[][202 models]{Magic:2013fd}. Also, spatial
and temporal averages of the full RHD models have been constructed; such models
facilitate calculation of large grids of synthetic spectra at a reasonable
computation cost thus allowing to apply them in problems of stellar abundance
analysis \citep[e.g.][]{2011SoPh..268..255C, Bergemann:2012jh,
2013A&A...557A...7T}.

\texttt{StaggerGrid project} \citep{Collet:2011ga} is a collaborative project
for the construction of a comprehensive grid of time-dependent 3D RHD model
atmospheres of solar- and late-type stars. The project involves several research
groups and different RHD codes.
\section{Line formation and spectrum synthesis}
\begin{figure}[b]
\begin{center}
\includegraphics[scale=.5]{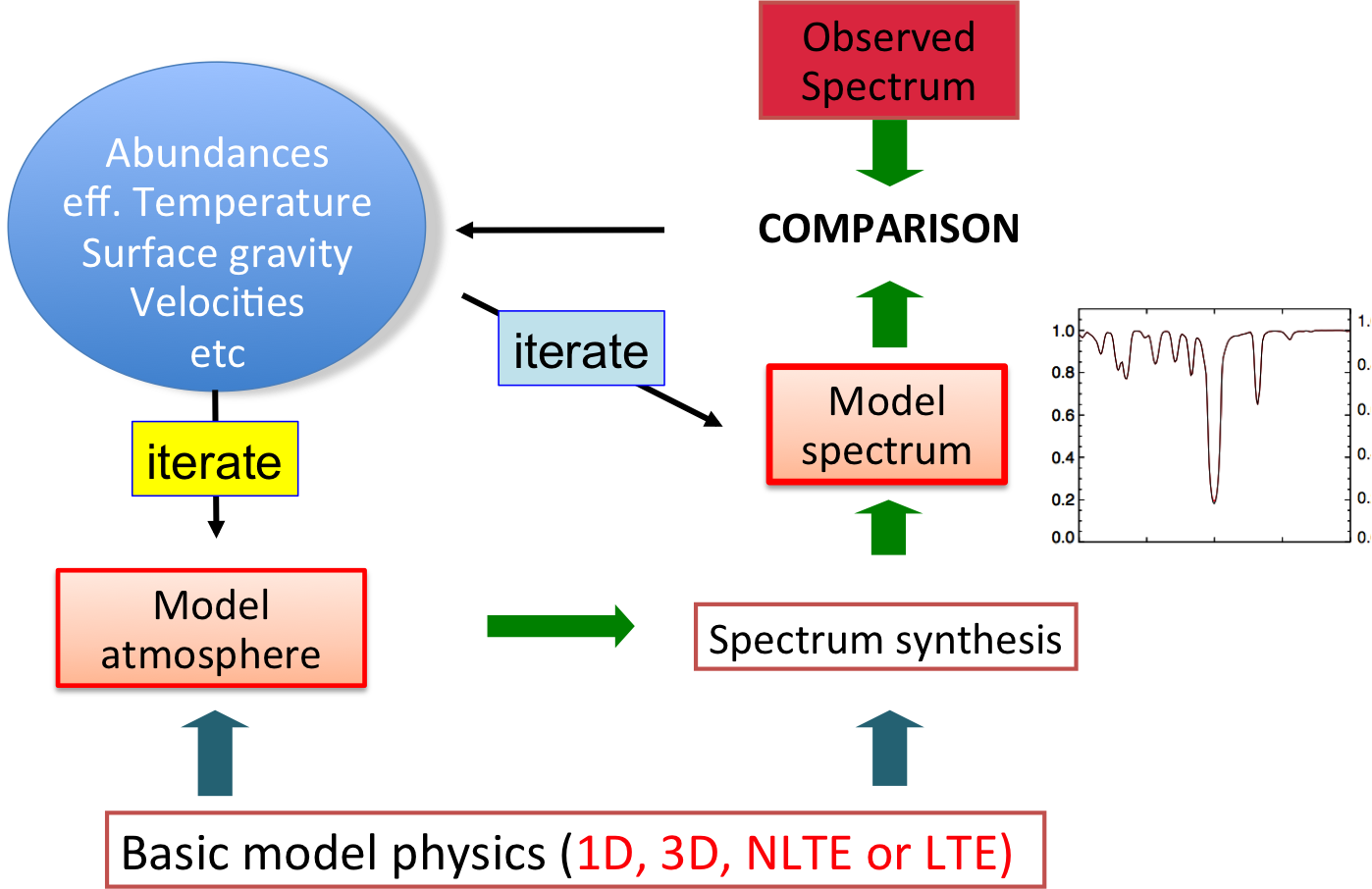}
\caption{The basic flow chart of spectroscopic analysis \label{fig:chart}.}
\end{center}
\end{figure}

Once we have a model providing basic thermodynamic quantities of an atmosphere
as a function of depth, it is straightforward to calculate the emergent stellar
spectra. The model spectra are then directly compared with observations and
stellar parameters can be determined from the best-fit templates (Fig.
\ref{fig:chart}).

\subsection{LTE spectrum synthesis}

In LTE, the calculation of model spectra can be easily done in 1D or 3D
geometry, by computing the formal solution of the radiative transfer equation
along a series of rays at different $\mu$ angles using the long- or
short-characteristics methods (see e.g. \citealt{Hauschildt:2006id}). Long
characteristics is preferred for
detailed line profiles, while short characteristics is used for solving
approximate radiation transport in 3D RH simulations. If scattering is included,
iterative solutions are needed, such as variable Eddington factor
\citep{Auer:1970wa}. 

In 1D LTE, very efficient and well-tested codes to compute full synthetic
spectra are SIU (\citealt{reetz}, \citealt{Schoenrich:2013uu}) and Turbospectrum
\citep{Alvarez:1998tg, 2012ascl.soft05004P}, SYNTHE \citep{Kurucz:2005vp}. These
codes were used to compute full libraries of synthetic spectra
\citep[e.g.][]{2004MNRAS.351.1430M, 2012A&A...544A.126D} for stellar
spectroscopy or population synthesis.

Codes that are able to compute 3D line formation in LTE include SCATE
\citep{Hayek:2011ft}, Optim3D \citep{Chiavassa:2009ck}, ASSET
\citep{Koesterke:2008dc}, Linfor3D \citep{2007A&A...473L..37C}. These codes can
potentially be used to compute full 3D grids of synthetic spectra. For example,
a grid of 3D synthetic spectra has been
recently presented by \citet{2013A&A...550A.103A}.

\begin{figure}[b]
\begin{center}\includegraphics[scale=.45,angle=-90]{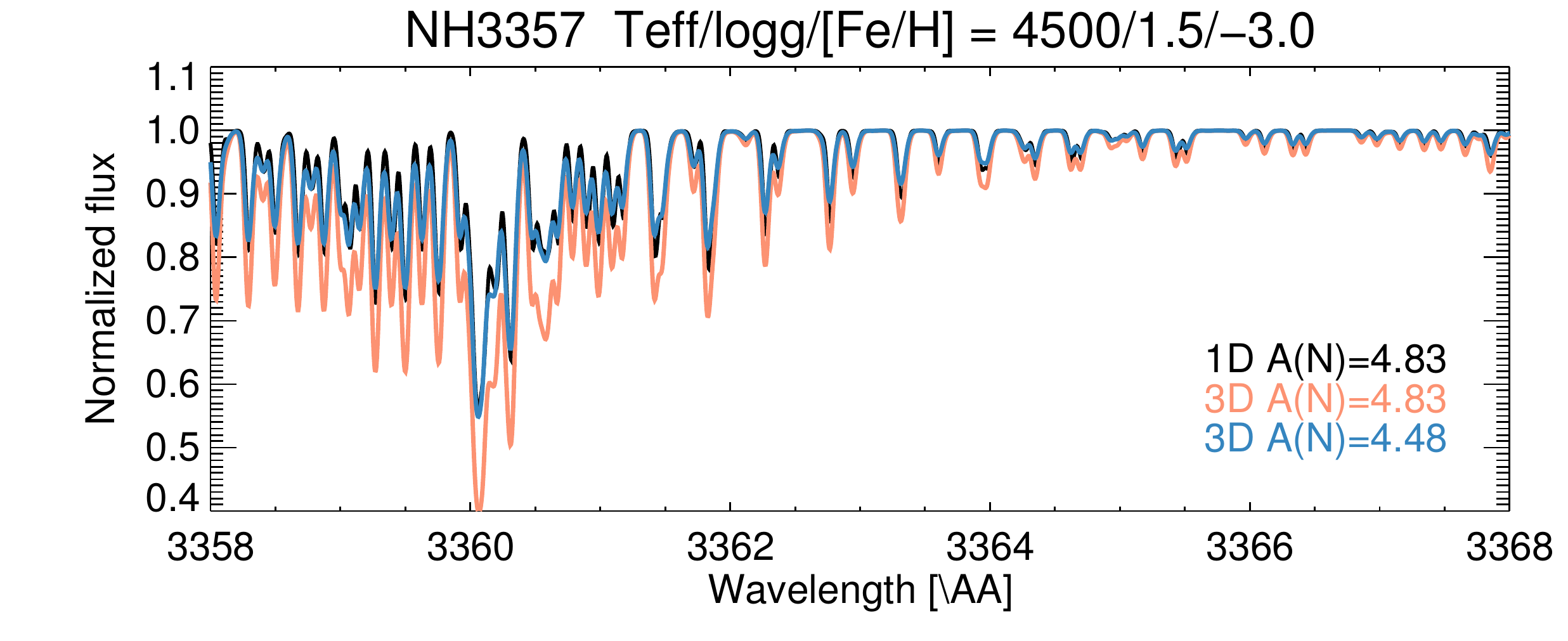}
\caption{CN lines computed with a 3D RHD model of a metal-poor star (Collet et
al. in prep). }\label{fig:cn}
\end{center}
\end{figure}

Fig. \ref{fig:cn} shows the CN lines computed with a 3D RHD model of a
metal-poor giant, in comparison with the 1D hydrostatic model (Collet in prep.).
In addition to a different temperature structure, temperature and pressure
fluctuations in the 3D models contribute strictly positively to the molecular
number density \citep{Uitenbroek:2011fe}, that is why the abundance derived from
molecular lines is lower. in Fig. \ref{fig:cn}, the CN lines in 3D provide a
$-0.4$ dex lower abundances compared to the 1D result.

\subsection{NLTE}
In NLTE, the problem of radiative transfer and thus modelling a spectrum is much
more complex. Radiative transfer shall be solved simultaneously with the rate
equations, the latter provide coupling between non-local radiation field and
local properties of the gas. 
\begin{equation}
\frac{\partial n_i}{\partial t} + \nabla \cdot (n_i \pmb{\upsilon}) =
\sum_{j \neq i} n_j P_{ji} - n_i \sum_{j \neq i} P_{ij}
\end{equation}
where $n$ is the number density of species of a certain type (e.g. atoms on the
excitation state $i$), $P$ the total transition rates (radiative plus
collisional) between the states. In the time-independent case, the left-hand
term can be neglected and we recover the statistical equilibrium (SE) equations.
Even in the 1D hydrostatic case, the solution of the RT and SE  requires very
efficient numerical schemes. In 1D, the most widely-used codes are DETAIL
(\citealt{Butler85} and the updated version by \citealt{2012ApJ...751..156B})
and MULTI \citep{1992ASPC...26..499C}. The codes are stable and efficient.
Presently, they are most useful for the diagnostics and
detailed analysis of NLTE effects, especially for complex multi-electron atomic
systems, like Fe I, and applications to stellar parameter and abundance analysis
of large datasets.

\begin{figure}[b]
\begin{center}
\includegraphics[scale=.55]{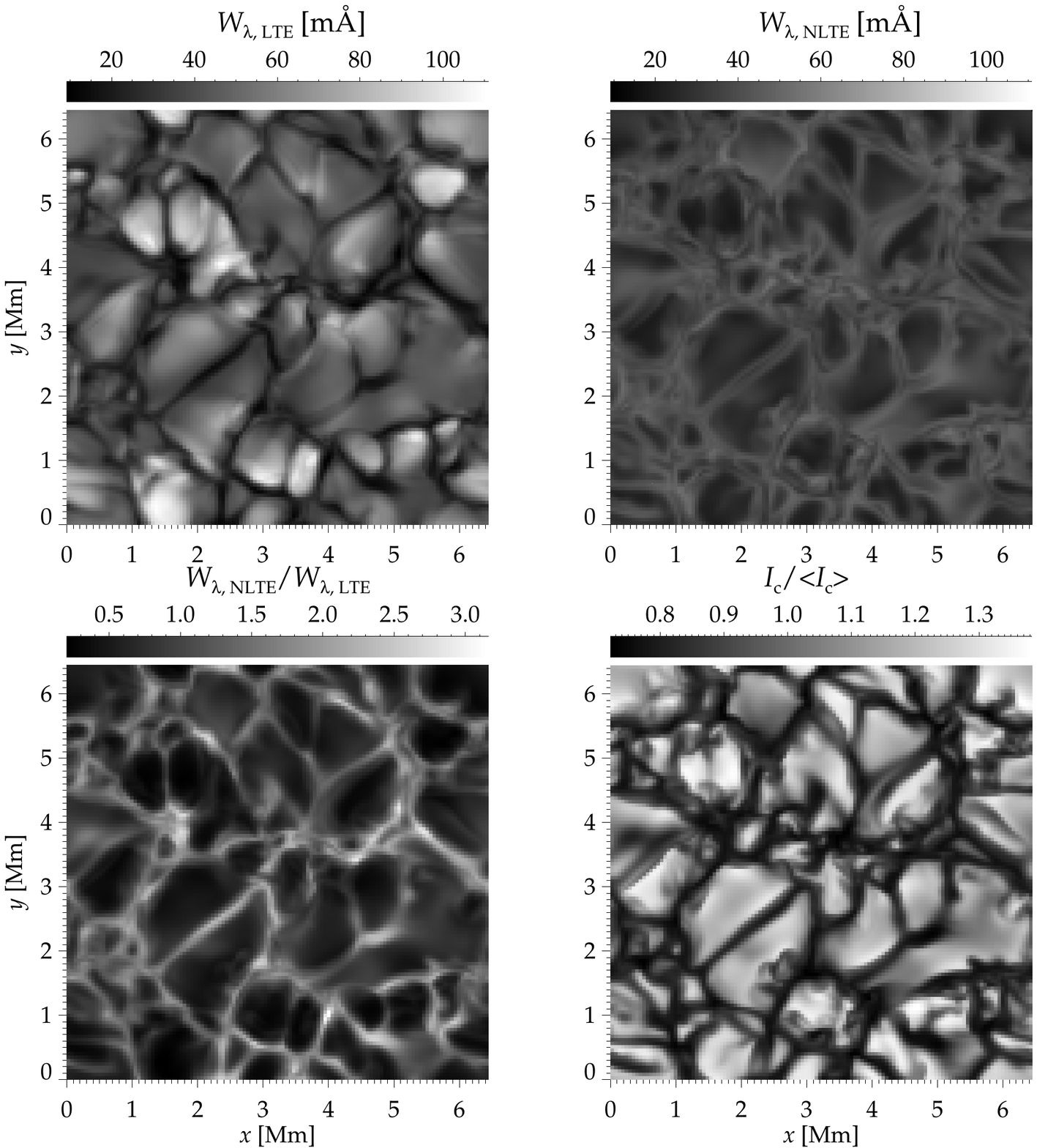}
\caption{Variation of the equivalent width for a Li I line at $6707$ \AA across
a surface of a metal-poor star; the NLTE radiative transfer was computed using
the 3D RHD simulation \citep{Lind:2013ge}. Reproduced from permission by the
author and the publisher.
\label{fig:li}}
\end{center}
\end{figure}

In 3D, MULTI3D \citep{1999ASSL..240..379B, 2009ASPC..415...87L} and MUGA
\citep{1994A&A...292..599A} are frequently used for the analysis of stellar
observations. The codes have been already applied to solve line transfer
problems in the Sun and late-type stars. For example, Fig. \ref{fig:li} shows
how the Li I line strength varies across the surface of star in a 3D
radiation-hydrodynamic simulation \citep{Lind:2013ge}. Other interesting
applications of the fully 3D NLTE framework include \citet[][Li
I]{2003A&A...399L..31A}, \citet[][Li I]{2007A&A...473L..37C}, \citet[][O
I]{2004A&A...417..751A}, \citet[][Sr I]{2007ApJ...664L.135T}, \citet[][Ca
II]{2009ApJ...694L.128L}, \cite[][Ca I, Na I]{Lind:2013ge}. It is now also
possible to perform NLTE line formation with the temporal and spatial averages
of 3D models, such as done by \citet{Bergemann:2012jh} for Fe I  and
\citet{2013EAS....60..103B} for Ti I and Si I lines.

In general, NLTE effects on spectral line formation are well-understood for most
of the chemical elements. The majority of species, for which spectral lines are
observed in the spectra of FGKM stars, are affected by NLTE over-ionization.
That means, the NLTE lines are weaker, especially if they are on the linear part
of the curve-of-growth. In this case, the NLTE effects on the ionisation balance
are typically large for giants and metal-poor stars
\citep[e.g.][]{Bergemann:2012jh}. Resonance line scattering is relevant for
strong lines of singly-ionised elements \citep[Ba II, Sr
II:][]{2012A&A...546A..90B, 2009A&A...506.1393S}. Atomic lines of neutral
species in the IR may show emission line cores \citep{1992A&A...259L..53C,
1992A&A...253..567C}.  A detailed discussion of the NLTE effects in the spectral
lines of different elements can be found in \citet{2013arXiv1306.6426M}.

While the detailed solution of NLTE radiative transfer in 1D is straightforward,
this is not always simple, especially when it concerns stellar abundance
analysis. The most difficult task is usually to assemble a comprehensive model
atom using the best atomic data available for a given element. This involves
mining large atomic databases, like TOPbase and Kurucz web-servers, NIST, VALD;
it is even more difficult to homogenise the atomic information, e.g. electronic
configurations and level energies, particularly when the combination of
quantum-mechanical and experimental data is needed to ensure the completeness of
the model atom. For those workers in the field of observational stellar
astrophysics, who are not directly involved in the development of NLTE radiative
transfer models, this is a difficult undertaking.

A simpler approach has been developed, which allows to bypass difficult
statistical equilibrium calculations. The idea is to use \textit{NLTE abundance
corrections}, which can be pre-computed for a large grid of model atmospheres
and element abundances. The NLTE abundance correction is then applied to a
best-fit LTE abundance. Numerous tests showed that the accuracy of NLTE element
abundances obtained by this method is very good.  
\begin{svgraybox}
NLTE abundance correction for a given chemical element,  $\Delta_{\rm El}$, is
defined as:
\begin{equation}
\Delta_{\rm El} = \log \rm{A (El)}_{\rm NLTE} - \log \rm{A (El)}_{\rm LTE}
\end {equation}
i.e., it is the the logarithmic correction, which has to be applied to an LTE
abundance determination A of a specific line to obtain the correct value
corresponding to the use of NLTE line formation. 
\end{svgraybox} 

\texttt{INSPECT}\footnote{www.inspect-stars.net} is the first project, which
allows one to compute NLTE corrections online; one may also request LTE or NLTE
abundance for a measured equivalent width of a spectral line, provided stellar
parameters are known. The following species are available: Li I, Na I, Ti I, Fe
I/II, Sr II. Smaller databases of NLTE corrections are also provided in the
literature (e.g., \citealt{2008A&A...492..823B}: Mn I, 
\citealt{2011MNRAS.418..863M}: Mg I, Ca I).

\section{Conclusions}
We are now entering the new era of observational stellar astrophysics, moving
away from 1D hydrostatic models with LTE - which is known as 'classical'
approach - to 3D radiative hydrodynamics with non-LTE. The transition is slow,
mostly because of the associated computational challenges. However, the need for
more realistic models, which provide more accurate and un-biased results, i.e.
fundamental stellar parameters and element abundances, is now as urgent as never
before. Large-scale stellar surveys (like Gaia-ESO and APOGEE) provide observed
spectra of unprecedented quality and pave the way to massive applications of the
new models. From the recent developments in theory, it seems that most promising
approach is to compute NLTE line formation with the averages of full 3D RHD
models. 
\begin{acknowledgement}
Figure 1 reproduced by permission of the authors; the image was
observed with the Swedish 1-m Solar Telescope. The SST is operated on
the island of La Palma by the Institute for Solar Physics in the
Spanish Observatorio del Roque de los Muchachos of the Instituto de
Astrofisica de Canarias. The Institute for Solar Physics is a national
research infrastructure under the Swedish Research Council. It is
managed as an independent institute associated with Stockholm
University through its Department of Astronomy. Figure 2: Asplund et
al. A\&A, 359, 729, 2000, reproduced with permission (c) ESO. Figures
3, 4 reproduced by permission of the AAS. Figure 8: Lind et al. A\&A,
554, A96, 2013, reproduced with permission (c) ESO. We thank R. Collet and R.
Trampedach for the figures from the papers in preparation, and K. Lind for the
useful comments to the manuscript. This work was partly supported by the European Union FP7 programme through ERC grant number 320360.
\end{acknowledgement}
\bibliographystyle{spbasic}
\bibliography{references}

\begin{thebibliography}{82}
\providecommand{\natexlab}[1]{#1}
\providecommand{\url}[1]{{#1}}
\providecommand{\urlprefix}{URL }
\expandafter\ifx\csname urlstyle\endcsname\relax
  \providecommand{\doi}[1]{DOI~\discretionary{}{}{}#1}\else
  \providecommand{\doi}{DOI~\discretionary{}{}{}\begingroup
  \urlstyle{rm}\Url}\fi
\providecommand{\eprint}[2][]{\url{#2}}

\bibitem[{{Allende Prieto} et~al(2013){Allende Prieto}, {Koesterke}, {Ludwig},
  {Freytag}, and {Caffau}}]{2013A&A...550A.103A}
{Allende Prieto} C, {Koesterke} L, {Ludwig} HG, {Freytag} B, {Caffau} E (2013)
  {Convective line shifts for the Gaia RVS from the CIFIST 3D model atmosphere
  grid}. \aap 550:A103, \doi{10.1051/0004-6361/201220064}, \eprint{1301.3703}

\bibitem[{Alvarez and Plez(1998)}]{Alvarez:1998tg}
Alvarez R, Plez B (1998) {Near-infrared narrow-band photometry of M-giant and
  Mira stars: models meet observations}. Astronomy {\&} Astrophysics
  330:1109--1119

\bibitem[{Asplund(2005)}]{2005ARA&A..43..481A}
Asplund M (2005) {New Light on Stellar Abundance Analyses: Departures from LTE
  and Homogeneity}. Annual Review of Astronomy and Astrophysics 43(1):481--530

\bibitem[{Asplund et~al(2000)Asplund, Nordlund, Trampedach, Allende~Prieto, and
  Stein}]{2000A&A...359..729A}
Asplund M, Nordlund A, Trampedach R, Allende~Prieto C, Stein RF (2000) {Line
  formation in solar granulation. I. Fe line shapes, shifts and asymmetries}.
  Astronomy {\&} Astrophysics 359:729--742

\bibitem[{{Asplund} et~al(2003){Asplund}, {Carlsson}, and
  {Botnen}}]{2003A&A...399L..31A}
{Asplund} M, {Carlsson} M, {Botnen} AV (2003) {Multi-level 3D non-LTE
  computations of lithium lines in the metal-poor halo stars HD 140283 and HD
  84937}. \aap 399:L31--L34, \doi{10.1051/0004-6361:20030080},
  \eprint{astro-ph/0302406}

\bibitem[{{Asplund} et~al(2004){Asplund}, {Grevesse}, {Sauval}, {Allende
  Prieto}, and {Kiselman}}]{2004A&A...417..751A}
{Asplund} M, {Grevesse} N, {Sauval} AJ, {Allende Prieto} C, {Kiselman} D (2004)
  {Line formation in solar granulation. IV. [O I], O I and OH lines and the
  photospheric O abundance}. \aap 417:751--768,
  \doi{10.1051/0004-6361:20034328}, \eprint{astro-ph/0312290}

\bibitem[{{Auer} et~al(1994){Auer}, {Bendicho}, and {Trujillo
  Bueno}}]{1994A&A...292..599A}
{Auer} L, {Bendicho} PF, {Trujillo Bueno} J (1994) {Multidimensional radiative
  transfer with multilevel atoms. 1: ALI method with preconditioning of the
  rate equations}. \aap 292:599--615

\bibitem[{Auer and Mihalas(1970)}]{Auer:1970wa}
Auer LH, Mihalas D (1970) {On the use of variable Eddington factors in non-LTE
  stellar atmospheres computations}. Monthly Notices of the Royal Astronomical
  Society 149:65

\bibitem[{Baron and Hauschildt(1998)}]{Baron:1998dn}
Baron E, Hauschildt PH (1998) {Parallel Implementation of the PHOENIX
  Generalized Stellar Atmosphere Program. II. Wavelength Parallelization}. The
  Astrophysical Journal 495(1):370--376

\bibitem[{Beeck et~al(2012)Beeck, Collet, Steffen, Asplund, Cameron, Freytag,
  Hayek, Ludwig, and Sch{\"u}ssler}]{Beeck:2012hn}
Beeck B, Collet R, Steffen M, Asplund M, Cameron RH, Freytag B, Hayek W, Ludwig
  HG, Sch{\"u}ssler M (2012) {Simulations of the solar near-surface layers with
  the CO5BOLD, MURaM, and Stagger codes}. Astronomy {\&} Astrophysics 539:A121

\bibitem[{{Bergemann} and {Gehren}(2008)}]{2008A&A...492..823B}
{Bergemann} M, {Gehren} T (2008) {NLTE abundances of Mn in a sample of
  metal-poor stars}. \aap 492:823--831, \doi{10.1051/0004-6361:200810098},
  \eprint{0811.0681}

\bibitem[{{Bergemann} et~al(2012{\natexlab{a}}){Bergemann}, {Hansen},
  {Bautista}, and {Ruchti}}]{2012A&A...546A..90B}
{Bergemann} M, {Hansen} CJ, {Bautista} M, {Ruchti} G (2012{\natexlab{a}}) {NLTE
  analysis of Sr lines in spectra of late-type stars with new R-matrix atomic
  data}. \aap 546:A90, \doi{10.1051/0004-6361/201219406}, \eprint{1207.2451}

\bibitem[{{Bergemann} et~al(2012{\natexlab{b}}){Bergemann}, {Kudritzki},
  {Plez}, {Davies}, {Lind}, and {Gazak}}]{2012ApJ...751..156B}
{Bergemann} M, {Kudritzki} RP, {Plez} B, {Davies} B, {Lind} K, {Gazak} Z
  (2012{\natexlab{b}}) {Red Supergiant Stars as Cosmic Abundance Probes: NLTE
  Effects in J-band Iron and Titanium Lines}. \apj 751:156,
  \doi{10.1088/0004-637X/751/2/156}, \eprint{1204.0511}

\bibitem[{Bergemann et~al(2012)Bergemann, Lind, Collet, Magic, and
  Asplund}]{Bergemann:2012jh}
Bergemann M, Lind K, Collet R, Magic Z, Asplund M (2012) {Non-LTE line
  formation of Fe in late-type stars -- I. Standard stars with 1D and〈3D〉
  model atmospheres}. Monthly Notices of the Royal Astronomical Society
  427(1):27--49

\bibitem[{{Bergemann} et~al(2013){Bergemann}, {Kudritzki}, {Davies}, {Plez},
  {Gazak}, and {Chiavassa}}]{2013EAS....60..103B}
{Bergemann} M, {Kudritzki} RP, {Davies} B, {Plez} B, {Gazak} Z, {Chiavassa} A
  (2013) {3D NLTE line formation in the atmospheres of red supergiants}. In:
  {Kervella} P, {Le Bertre} T, {Perrin} G (eds) EAS Publications Series, EAS
  Publications Series, vol~60, pp 103--109, \doi{10.1051/eas/1360011},
  \eprint{1303.4768}

\bibitem[{B{\"o}hm-Vitense(1958)}]{BohmVitense:1958vy}
B{\"o}hm-Vitense E (1958) {{\"U}ber die Wasserstoffkonvektionszone in Sternen
  verschiedener Effektivtemperaturen und Leuchtkr{\"a}fte. Mit 5
  Textabbildungen}. Zeitschrift fur Astrophysik 46:108

\bibitem[{{Botnen} and {Carlsson}(1999)}]{1999ASSL..240..379B}
{Botnen} A, {Carlsson} M (1999) {Multi3D, 3D Non-LTE Radiative Transfer}. In:
  {Miyama} SM, {Tomisaka} K, {Hanawa} T (eds) Numerical Astrophysics,
  Astrophysics and Space Science Library, vol 240, p 379

\bibitem[{{Buscher} et~al(1990){Buscher}, {Baldwin}, {Warner}, and
  {Haniff}}]{1990MNRAS.245P...7B}
{Buscher} DF, {Baldwin} JE, {Warner} PJ, {Haniff} CA (1990) {Detection of a
  bright feature on the surface of Betelgeuse}. \mnras 245:7P--11P

\bibitem[{{Butler} and {Giddings}(1985)}]{Butler85}
{Butler} K, {Giddings} J (1985) Newsletter on Analysis of Astronomical Spectra,
  University of London 9

\bibitem[{{Caffau} et~al(2011){Caffau}, {Ludwig}, {Steffen}, {Freytag}, and
  {Bonifacio}}]{2011SoPh..268..255C}
{Caffau} E, {Ludwig} HG, {Steffen} M, {Freytag} B, {Bonifacio} P (2011) {Solar
  Chemical Abundances Determined with a CO5BOLD 3D Model Atmosphere}. \solphys
  268:255--269, \doi{10.1007/s11207-010-9541-4}, \eprint{1003.1190}

\bibitem[{Canuto and Mazzitelli(1991)}]{Canuto:1991bj}
Canuto VM, Mazzitelli I (1991) {Stellar turbulent convection - A new model and
  applications}. The Astrophysical Journal 370:295--311

\bibitem[{{Carlsson}(1992)}]{1992ASPC...26..499C}
{Carlsson} M (1992) {The MULTI Non-LTE Program (Invited Review)}. In:
  {Giampapa} MS, {Bookbinder} JA (eds) Cool Stars, Stellar Systems, and the
  Sun, Astronomical Society of the Pacific Conference Series, vol~26, p 499

\bibitem[{Carlsson and Rutten(1992)}]{1992A&A...259L..53C}
Carlsson M, Rutten RJ (1992) {Solar hydrogen lines in the infrared}. Astronomy
  and Astrophysics (ISSN 0004-6361) 259:L53--L56

\bibitem[{Carlsson et~al(1992)Carlsson, Rutten, and
  Shchukina}]{1992A&A...253..567C}
Carlsson M, Rutten RJ, Shchukina NG (1992) {The formation of the MG I emission
  features near 12 microns}. Astronomy and Astrophysics (ISSN 0004-6361)
  253:567--585

\bibitem[{{Cayrel} et~al(2007){Cayrel}, {Steffen}, {Chand}, {Bonifacio},
  {Spite}, {Spite}, {Petitjean}, {Ludwig}, and {Caffau}}]{2007A&A...473L..37C}
{Cayrel} R, {Steffen} M, {Chand} H, {Bonifacio} P, {Spite} M, {Spite} F,
  {Petitjean} P, {Ludwig} HG, {Caffau} E (2007) {Line shift, line asymmetry,
  and the 6Li/7Li isotopic ratio determination}. \aap 473:L37--L40,
  \doi{10.1051/0004-6361:20078342}, \eprint{0708.3819}

\bibitem[{Chiavassa et~al(2009)Chiavassa, Plez, Josselin, and
  Freytag}]{Chiavassa:2009ck}
Chiavassa A, Plez B, Josselin E, Freytag B (2009) {Radiative hydrodynamics
  simulations of red supergiant stars}. Astronomy {\&} Astrophysics
  506(3):1351--1365

\bibitem[{Chiavassa et~al(2011)Chiavassa, Freytag, Masseron, and
  Plez}]{Chiavassa:2011ew}
Chiavassa A, Freytag B, Masseron T, Plez B (2011) {Radiative hydrodynamics
  simulations of red supergiant stars}. Astronomy {\&} Astrophysics 535:A22

\bibitem[{{Collet} et~al(2007){Collet}, {Asplund}, and
  {Trampedach}}]{2007A&A...469..687C}
{Collet} R, {Asplund} M, {Trampedach} R (2007) {Three-dimensional
  hydrodynamical simulations of surface convection in red giant stars. Impact
  on spectral line formation and abundance analysis}. \aap 469:687--706,
  \doi{10.1051/0004-6361:20066321}, \eprint{arXiv:astro-ph/0703652}

\bibitem[{Collet et~al(2011)Collet, Magic, and Asplund}]{Collet:2011ga}
Collet R, Magic Z, Asplund M (2011) {The StaggerGrid project: a grid of 3-D
  model atmospheres for high-precision spectroscopy}. Journal of Physics:
  Conference Series 328(1):012,003

\bibitem[{{de Laverny} et~al(2012){de Laverny}, {Recio-Blanco}, {Worley}, and
  {Plez}}]{2012A&A...544A.126D}
{de Laverny} P, {Recio-Blanco} A, {Worley} CC, {Plez} B (2012) {The AMBRE
  project: A new synthetic grid of high-resolution FGKM stellar spectra}. \aap
  544:A126, \doi{10.1051/0004-6361/201219330}, \eprint{1205.2270}

\bibitem[{Freytag et~al(2002)Freytag, Steffen, and Dorch}]{2002AN....323..213F}
Freytag B, Steffen M, Dorch B (2002) {Spots on the surface of Betelgeuse --
  Results from new 3D stellar convection models}. Astronomische Nachrichten
  323:213--219

\bibitem[{{Freytag} et~al(2012){Freytag}, {Steffen}, {Ludwig},
  {Wedemeyer-B{\"o}hm}, {Schaffenberger}, and {Steiner}}]{2012JCoPh.231..919F}
{Freytag} B, {Steffen} M, {Ludwig} HG, {Wedemeyer-B{\"o}hm} S, {Schaffenberger}
  W, {Steiner} O (2012) {Simulations of stellar convection with CO5BOLD}.
  Journal of Computational Physics 231:919--959,
  \doi{10.1016/j.jcp.2011.09.026}, \eprint{1110.6844}

\bibitem[{{Gray} et~al(2008){Gray}, {Carney}, and {Yong}}]{2008AJ....135.2033G}
{Gray} DF, {Carney} BW, {Yong} D (2008) {Asymmetries in the Spectral Lines of
  Evolved Halo Stars}. \aj 135:2033--2037, \doi{10.1088/0004-6256/135/6/2033}

\bibitem[{Grupp(2004)}]{Grupp:2004jc}
Grupp F (2004) {MAFAGS-OS: New opacity sampling model atmospheres for A, F and
  G stars}. Astronomy {\&} Astrophysics 426(1):309--322

\bibitem[{{Gudiksen} et~al(2011){Gudiksen}, {Carlsson}, {Hansteen}, {Hayek},
  {Leenaarts}, and {Mart{\'{\i}}nez-Sykora}}]{2011A&A...531A.154G}
{Gudiksen} BV, {Carlsson} M, {Hansteen} VH, {Hayek} W, {Leenaarts} J,
  {Mart{\'{\i}}nez-Sykora} J (2011) {The stellar atmosphere simulation code
  Bifrost. Code description and validation}. \aap 531:A154,
  \doi{10.1051/0004-6361/201116520}

\bibitem[{Gustafsson et~al(2008)Gustafsson, Edvardsson, Eriksson, J{\o}rgensen,
  Nordlund, and Plez}]{Gustafsson:2008df}
Gustafsson B, Edvardsson B, Eriksson K, J{\o}rgensen UG, Nordlund A, Plez B
  (2008) {A grid of MARCS model atmospheres for late-type stars}. Astronomy
  {\&} Astrophysics 486(3):951--970

\bibitem[{Hansteen et~al(2007)Hansteen, Carlsson, and
  Gudiksen}]{Hansteen:2007wn}
Hansteen VH, Carlsson M, Gudiksen B (2007) {3D Numerical Models of the
  Chromosphere, Transition Region, and Corona}. The Physics of Chromospheric
  Plasmas 368:107

\bibitem[{{Haubois} et~al(2009){Haubois}, {Perrin}, {Lacour}, {Verhoelst},
  {Meimon}, {Mugnier}, {Thi{\'e}baut}, {Berger}, {Ridgway}, {Monnier},
  {Millan-Gabet}, and {Traub}}]{2009A&A...508..923H}
{Haubois} X, {Perrin} G, {Lacour} S, {Verhoelst} T, {Meimon} S, {Mugnier} L,
  {Thi{\'e}baut} E, {Berger} JP, {Ridgway} ST, {Monnier} JD, {Millan-Gabet} R,
  {Traub} W (2009) {Imaging the spotty surface of Betelgeuse in the H band}.
  \aap 508:923--932, \doi{10.1051/0004-6361/200912927}, \eprint{0910.4167}

\bibitem[{Hauschildt and Baron(2006)}]{Hauschildt:2006id}
Hauschildt PH, Baron E (2006) {A 3D radiative transfer framework}. Astronomy
  {\&} Astrophysics 451(1):273--284

\bibitem[{Hayek et~al(2011)Hayek, Asplund, Collet, and Nordlund}]{Hayek:2011ft}
Hayek W, Asplund M, Collet R, Nordlund A (2011) {3D LTE spectral line formation
  with scattering in red giant stars}. Astronomy {\&} Astrophysics 529:A158

\bibitem[{H{\"o}fner et~al(2003)H{\"o}fner, Gautschy-Loidl, Aringer, Andersen,
  and J{\o}rgensen}]{Hofner:2003vf}
H{\"o}fner S, Gautschy-Loidl R, Aringer B, Andersen A, J{\o}rgensen UG (2003)
  {Dynamical Atmospheres and Winds of AGB Stars}. Astronomische Nachrichten
  Supplement 324:19

\bibitem[{Kervella et~al(2009)Kervella, Verhoelst, Ridgway, Perrin, Lacour,
  Cami, and Haubois}]{Kervella:2009di}
Kervella P, Verhoelst T, Ridgway ST, Perrin G, Lacour S, Cami J, Haubois X
  (2009) {The close circumstellar environment of Betelgeuse}. Astronomy {\&}
  Astrophysics 504(1):115--125

\bibitem[{Koesterke et~al(2008)Koesterke, Allende~Prieto, and
  Lambert}]{Koesterke:2008dc}
Koesterke L, Allende~Prieto C, Lambert DL (2008) {Center‐to‐Limb Variation
  of Solar Three‐dimensional Hydrodynamical Simulations}. The Astrophysical
  Journal 680(1):764--773

\bibitem[{Kurucz(1993)}]{Kurucz:1993wc}
Kurucz RL (1993) {A New Opacity-Sampling Model Atmosphere Program for Arbitrary
  Abundances}. The Physics of Chromospheric Plasmas 44:87

\bibitem[{Kurucz(2005)}]{Kurucz:2005vp}
Kurucz RL (2005) {ATLAS12, SYNTHE, ATLAS9, WIDTH9, et cetera}. Memorie della
  Societa Astronomica Italiana 8:14

\bibitem[{{Leenaarts} and {Carlsson}(2009)}]{2009ASPC..415...87L}
{Leenaarts} J, {Carlsson} M (2009) {MULTI3D: A Domain-Decomposed 3D Radiative
  Transfer Code}. In: {Lites} B, {Cheung} M, {Magara} T, {Mariska} J, {Reeves}
  K (eds) The Second Hinode Science Meeting: Beyond Discovery-Toward
  Understanding, Astronomical Society of the Pacific Conference Series, vol
  415, p~87

\bibitem[{{Leenaarts} et~al(2009){Leenaarts}, {Carlsson}, {Hansteen}, and
  {Rouppe van der Voort}}]{2009ApJ...694L.128L}
{Leenaarts} J, {Carlsson} M, {Hansteen} V, {Rouppe van der Voort} L (2009)
  {Three-Dimensional Non-LTE Radiative Transfer Computation of the CA 8542
  Infrared Line From a Radiation-MHD Simulation}. \apjl 694:L128--L131,
  \doi{10.1088/0004-637X/694/2/L128}, \eprint{0903.0791}

\bibitem[{Lind et~al(2013)Lind, Melendez, Asplund, Collet, and
  Magic}]{Lind:2013ge}
Lind K, Melendez J, Asplund M, Collet R, Magic Z (2013) {The lithium isotopic
  ratio in very metal-poor stars}. Astronomy {\&} Astrophysics 554:A96

\bibitem[{Lucy(1999)}]{Lucy:1999wd}
Lucy LB (1999) {Computing radiative equilibria with Monte Carlo techniques}.
  Astronomy {\&} Astrophysics 344:282--288

\bibitem[{{Ludwig} et~al(2009){Ludwig}, {Caffau}, {Steffen}, {Freytag},
  {Bonifacio}, and {Ku{\v c}inskas}}]{2009MmSAI..80..711L}
{Ludwig} HG, {Caffau} E, {Steffen} M, {Freytag} B, {Bonifacio} P, {Ku{\v
  c}inskas} A (2009) {The CIFIST 3D model atmosphere grid.} \memsai 80:711,
  \eprint{0908.4496}

\bibitem[{Magic et~al(2013)Magic, Collet, Asplund, Trampedach, Hayek,
  Chiavassa, Stein, and Nordlund}]{Magic:2013fd}
Magic Z, Collet R, Asplund M, Trampedach R, Hayek W, Chiavassa A, Stein RF,
  Nordlund A (2013) {The Stagger-grid: A grid of 3D stellar atmosphere models}.
  Astronomy {\&} Astrophysics 557:A26

\bibitem[{{Mashonkina}(2013)}]{2013arXiv1306.6426M}
{Mashonkina} L (2013) {Review: progress in NLTE calculations and their
  application to large data-sets}. ArXiv e-prints \eprint{1306.6426}

\bibitem[{{Merle} et~al(2011){Merle}, {Th{\'e}venin}, {Pichon}, and
  {Bigot}}]{2011MNRAS.418..863M}
{Merle} T, {Th{\'e}venin} F, {Pichon} B, {Bigot} L (2011) {A grid of non-local
  thermodynamic equilibrium corrections for magnesium and calcium in late-type
  giant and supergiant stars: application to Gaia}. \mnras 418:863--887,
  \doi{10.1111/j.1365-2966.2011.19540.x}, \eprint{1107.6015}

\bibitem[{Mihalas(1979)}]{Mihalas:1979ux}
Mihalas D (1979) {Book-Review - Stellar Atmospheres}. Soviet Astronomy 23:386

\bibitem[{Mihalas and Athay(1973)}]{Mihalas:1973bu}
Mihalas D, Athay RG (1973) {The Effects of Departures from LTE in Stellar
  Spectra}. Annual Review of Astronomy and Astrophysics 11(1):187--218

\bibitem[{{Murphy} and {Meiksin}(2004)}]{2004MNRAS.351.1430M}
{Murphy} T, {Meiksin} A (2004) {A library of high-resolution Kurucz spectra in
  the range {$\lambda$}{$\lambda$}3000-10 000}. \mnras 351:1430--1438,
  \doi{10.1111/j.1365-2966.2004.07895.x}, \eprint{astro-ph/0404010}

\bibitem[{Muthsam et~al(2010)Muthsam, Kupka, L{\"o}w-Baselli, Obertscheider,
  Langer, and Lenz}]{Muthsam:2010dq}
Muthsam HJ, Kupka F, L{\"o}w-Baselli B, Obertscheider C, Langer M, Lenz P
  (2010) {ANTARES -- A Numerical Tool for Astrophysical RESearch with
  applications to solar granulation}. New Astronomy 15(5):460--475

\bibitem[{{Nordlund}(1982)}]{1982A&A...107....1N}
{Nordlund} A (1982) {Numerical simulations of the solar granulation. I - Basic
  equations and methods}. \aap 107:1--10

\bibitem[{Nordlund and Galsgaard(1992)}]{Nordlund:1992cq}
Nordlund {\AA}, Galsgaard K (1992) {Large scale simulations}. In:
  Electromechanical Coupling of the Solar Atmosphere, AIP, pp 13--23

\bibitem[{Nordlund et~al(2009)Nordlund, Stein, and Asplund}]{Nordlund:2009jf}
Nordlund {\AA}, Stein RF, Asplund M (2009) {Solar Surface Convection}. Living
  Reviews in Solar Physics 6:2

\bibitem[{{Ohnaka} et~al(2009){Ohnaka}, {Hofmann}, {Benisty}, {Chelli},
  {Driebe}, {Millour}, {Petrov}, {Schertl}, {Stee}, {Vakili}, and
  {Weigelt}}]{2009A&A...503..183O}
{Ohnaka} K, {Hofmann} KH, {Benisty} M, {Chelli} A, {Driebe} T, {Millour} F,
  {Petrov} R, {Schertl} D, {Stee} P, {Vakili} F, {Weigelt} G (2009) {Spatially
  resolving the inhomogeneous structure of the dynamical atmosphere of
  Betelgeuse with VLTI/AMBER}. \aap 503:183--195,
  \doi{10.1051/0004-6361/200912247}, \eprint{0906.4792}

\bibitem[{{Ohnaka} et~al(2011){Ohnaka}, {Weigelt}, {Millour}, {Hofmann},
  {Driebe}, {Schertl}, {Chelli}, {Massi}, {Petrov}, and
  {Stee}}]{2011A&A...529A.163O}
{Ohnaka} K, {Weigelt} G, {Millour} F, {Hofmann} KH, {Driebe} T, {Schertl} D,
  {Chelli} A, {Massi} F, {Petrov} R, {Stee} P (2011) {Imaging the dynamical
  atmosphere of the red supergiant Betelgeuse in the CO first overtone lines
  with VLTI/AMBER}. \aap 529:A163, \doi{10.1051/0004-6361/201016279},
  \eprint{1104.0958}

\bibitem[{Paxton et~al(2011)Paxton, Bildsten, Dotter, Herwig, Lesaffre, and
  Timmes}]{paxton:MESA}
Paxton B, Bildsten L, Dotter A, Herwig F, Lesaffre P, Timmes F (2011) Modules
  for experiments in stellar astrophysics {(MESA)}. ApJS 192(1):3:1--35

\bibitem[{{Plez}(2012)}]{2012ascl.soft05004P}
{Plez} B (2012) {Turbospectrum: Code for spectral synthesis}. Astrophysics
  Source Code Library, \eprint{1205.004}

\bibitem[{{Reetz}(1999)}]{reetz}
{Reetz} J (1999) {Oxygen Abundances in Cool Stars and the Chemical Evolution of
  the Galaxy}

\bibitem[{Schoenrich and Bergemann(2013)}]{Schoenrich:2013uu}
Schoenrich R, Bergemann M (2013) {Fundamental stellar parameters and
  metallicities from Bayesian spectroscopy. Application to low- and
  high-resolution spectra}. eprint arXiv:13115558

\bibitem[{{Shchukina} et~al(2009){Shchukina}, {Olshevsky}, and
  {Khomenko}}]{2009A&A...506.1393S}
{Shchukina} NG, {Olshevsky} VL, {Khomenko} EV (2009) {The solar Ba II 4554
  {\AA} line as a Doppler diagnostic: NLTE analysis in 3D hydrodynamical
  model}. \aap 506:1393--1404, \doi{10.1051/0004-6361/200912048},
  \eprint{0905.0985}

\bibitem[{Short and Hauschildt(2005)}]{Short:2005kh}
Short CI, Hauschildt PH (2005) {A Non‐LTE Line‐Blanketed Model of a
  Solar‐Type Star}. The Astrophysical Journal 618(2):926--938

\bibitem[{Short and Hauschildt(2009)}]{Short:2009iq}
Short CI, Hauschildt PH (2009) {NON-LTE MODELING OF THE NEAR-ULTRAVIOLET BAND
  OF LATE-TYPE STARS}. The Astrophysical Journal 691(2):1634--1647

\bibitem[{Spruit et~al(1990)Spruit, Nordlund, and Title}]{Spruit:1990cv}
Spruit HC, Nordlund A, Title AM (1990) {Solar Convection}. Annual Review of
  Astronomy and Astrophysics 28(1):263--303

\bibitem[{Stein and Nordlund(1998)}]{1998ApJ...499..914S}
Stein RF, Nordlund A (1998) {Simulations of Solar Granulation. I. General
  Properties}. Astrophysical Journal v499 499:914

\bibitem[{Tanner et~al(2013)Tanner, Basu, and Demarque}]{Tanner:2013gx}
Tanner JD, Basu S, Demarque P (2013) {VARIATION OF STELLAR ENVELOPE CONVECTION
  AND OVERSHOOT WITH METALLICITY}. The Astrophysical Journal 767(1):78

\bibitem[{Teyssier et~al(2012)Teyssier, Quintana-Lacaci, Marston, Bujarrabal,
  Alcolea, Cernicharo, Decin, Dominik, Justtanont, de~Koter, Melnick, Menten,
  Neufeld, Olofsson, Planesas, Schmidt, Soria-Ruiz, Sch{\"o}ier, Szczerba, and
  Waters}]{Teyssier:2012bz}
Teyssier D, Quintana-Lacaci G, Marston AP, Bujarrabal V, Alcolea J, Cernicharo
  J, Decin L, Dominik C, Justtanont K, de~Koter A, Melnick G, Menten KM,
  Neufeld DA, Olofsson H, Planesas P, Schmidt M, Soria-Ruiz R, Sch{\"o}ier FL,
  Szczerba R, Waters LBFM (2012) {Herschel/HIFI observations of red supergiants
  and yellow hypergiants}. Astronomy {\&} Astrophysics 545:A99

\bibitem[{{Trampedach} et~al(2013){Trampedach}, {Asplund}, {Collet},
  {Nordlund}, and {Stein}}]{2013ApJ...769...18T}
{Trampedach} R, {Asplund} M, {Collet} R, {Nordlund} {\AA}, {Stein} RF (2013) {A
  Grid of Three-dimensional Stellar Atmosphere Models of Solar Metallicity. I.
  General Properties, Granulation, and Atmospheric Expansion}. \apj 769:18,
  \doi{10.1088/0004-637X/769/1/18}, \eprint{1303.1780}

\bibitem[{{Tremblay} et~al(2013){Tremblay}, {Ludwig}, {Freytag}, {Steffen}, and
  {Caffau}}]{2013A&A...557A...7T}
{Tremblay} PE, {Ludwig} HG, {Freytag} B, {Steffen} M, {Caffau} E (2013)
  {Granulation properties of giants, dwarfs, and white dwarfs from the CIFIST
  3D model atmosphere grid}. \aap 557:A7, \doi{10.1051/0004-6361/201321878},
  \eprint{1307.2810}

\bibitem[{{Trujillo Bueno} and {Shchukina}(2007)}]{2007ApJ...664L.135T}
{Trujillo Bueno} J, {Shchukina} N (2007) {The Scattering Polarization of the Sr
  I {$\lambda$}4607 Line at the Diffraction Limit Resolution of a 1 m
  Telescope}. \apjl 664:L135--L138, \doi{10.1086/520838}, \eprint{0706.2386}

\bibitem[{{Tuthill} et~al(1997){Tuthill}, {Haniff}, and
  {Baldwin}}]{1997MNRAS.285..529T}
{Tuthill} PG, {Haniff} CA, {Baldwin} JE (1997) {Hotspots on late-type
  supergiants}. \mnras 285:529--539

\bibitem[{Uitenbroek and Criscuoli(2011)}]{Uitenbroek:2011fe}
Uitenbroek H, Criscuoli S (2011) {WHY ONE-DIMENSIONAL MODELS FAIL IN THE
  DIAGNOSIS OF AVERAGE SPECTRA FROM INHOMOGENEOUS STELLAR ATMOSPHERES}. The
  Astrophysical Journal 736(1):69

\bibitem[{{V{\"o}gler} et~al(2005){V{\"o}gler}, {Shelyag}, {Sch{\"u}ssler},
  {Cattaneo}, {Emonet}, and {Linde}}]{2005A&A...429..335V}
{V{\"o}gler} A, {Shelyag} S, {Sch{\"u}ssler} M, {Cattaneo} F, {Emonet} T,
  {Linde} T (2005) {Simulations of magneto-convection in the solar photosphere.
  Equations, methods, and results of the MURaM code}. \aap 429:335--351,
  \doi{10.1051/0004-6361:20041507}

\bibitem[{{Wilson} et~al(1992){Wilson}, {Baldwin}, {Buscher}, and
  {Warner}}]{1992MNRAS.257..369W}
{Wilson} RW, {Baldwin} JE, {Buscher} DF, {Warner} PJ (1992) {High-resolution
  imaging of Betelgeuse and Mira}. \mnras 257:369--376

\bibitem[{{Wilson} et~al(1997){Wilson}, {Dhillon}, and
  {Haniff}}]{1997MNRAS.291..819W}
{Wilson} RW, {Dhillon} VS, {Haniff} CA (1997) {The changing face of
  Betelgeuse}. \mnras 291:819

\bibitem[{{Young} et~al(2000){Young}, {Baldwin}, {Boysen}, {Haniff}, {Lawson},
  {Mackay}, {Pearson}, {Rogers}, {St.-Jacques}, {Warner}, {Wilson}, and
  {Wilson}}]{2000MNRAS.315..635Y}
{Young} JS, {Baldwin} JE, {Boysen} RC, {Haniff} CA, {Lawson} PR, {Mackay} CD,
  {Pearson} D, {Rogers} J, {St-Jacques} D, {Warner} PJ, {Wilson} DMA, {Wilson}
  RW (2000) {New views of Betelgeuse: multi-wavelength surface imaging and
  implications for models of hotspot generation}. \mnras 315:635--645,
  \doi{10.1046/j.1365-8711.2000.03438.x}

\end{thebibliography}
\end{document}